\documentclass[journal]{IEEEtran}
\usepackage[T1]{fontenc}

\usepackage{graphicx}
\usepackage[noadjust]{cite}
\usepackage{mcite}
\usepackage{amsfonts,helvet}
\usepackage{fancyhdr}
\usepackage{threeparttable}
\usepackage{epsf,epsfig}
\usepackage{amsthm}
\usepackage{amsmath}
\usepackage{siunitx}
\usepackage{amssymb}
\usepackage{dsfont}
\usepackage{subfigure}
\usepackage{color}
\usepackage{enumerate}
\usepackage{hyperref}
\usepackage{cancel}
\usepackage{bbm}
\usepackage{dsfont}
\usepackage[subnum]{cases}
\usepackage{adjustbox}
\usepackage[linesnumbered,ruled]{algorithm2e}
\usepackage{multicol}
\usepackage[english]{babel}
\usepackage{enumitem}
\usepackage{array}
\usepackage{etoolbox}
\usepackage[margin= 66.5 pt]{geometry}

\newmuskip\pFqmuskip

\makeatletter
\newcommand{\thickhline}{%
    \noalign {\ifnum 0=`}\fi \hrule height 1pt
    \futurelet \reserved@a \@xhline
}
\newcolumntype{"}{@{\hskip\tabcolsep\vrule width 1pt\hskip\tabcolsep}}
\makeatother

\IEEEoverridecommandlockouts

\title{
\huge Advanced Receiver Architectures for Millimeter Wave Communications with Low-Resolution ADCs
}
\author{
Jinseok Choi, {\it Member, IEEE}, Gilwon Lee, {\it Member, IEEE}, Ahmed Alkhateeb, {\it Member, IEEE}, \\Alan Gatherer, {\it Fellow, IEEE}, Naofal Al-Dhahir, {\it Fellow, IEEE}, and Brian L. Evans, {\it Fellow, IEEE}
}
\begin{document}
\maketitle

\begin{abstract}
\color{black}Employing low-resolution analog-to-digital converters (ADCs) for millimeter wave receivers with large antenna arrays provides opportunity to efficiently reduce power consumption of the receiver. 
Reducing ADC resolution, however, results in performance degradation due to non-negligible quantization error.
In addition, the large number of radio frequency (RF) chains is still not desirable.
Accordingly, conventional low-resolution ADC systems require more efficient designs to minimize the cost and complexity while maximizing performance.
In this article, we discuss advanced low-resolution ADC receiver architectures that further improve the spectral and energy efficiency tradeoff.
{\color{black}
To reduce both the numbers of RF chains and ADC bits, hybrid analog-and-digital beamforming is jointly considered with low-resolution ADCs.
We explore the challenges in designing such receivers
and present key insights on how the advanced architectures overcome such challenges. 
As an alternative low-resolution ADC receiver, we also introduce receivers with learning-based detection.
The receiver does not require explicit channel estimation, thereby is suitable for one-bit ADC systems.
Finally, future challenges and research issues are discussed.}
\end{abstract}





\section{Introduction}
\label{sec:intro}


Millimeter wave (mmWave) massive multiple-input multiple-output (MIMO) communication systems offer a substantial increase in data rate thanks to the availability of wide bandwidth and packing of many antennas in a small array space~\cite{pi2011introduction}. 
The problem of prohibitively large power consumption, however, becomes a major implementation challenge due to the large number of antennas coupled with high sampling rates.
Traditionally, the large antenna arrays accompany a large number of RF chains followed by a large number of high-resolution ADCs.
The power consumption of ADCs scales exponentially in the number of quantization bits $b$, thereby driving high-speed and high-resolution ADCs to be the primary power consumers in the mmWave massive MIMO systems~\cite{zhang2018low}. 


Accordingly, low-resolution ADC systems have attracted considerable attention as a low-power solution and showed improved tradeoff between spectral efficiency (SE) and energy efficiency (EE)~\cite{liu2019low}.
{\color{black} Nevertheless, 
there are certain limitations in using low-resolution ADCs, namely: (i) the SE loss from the non-negligible quantization error is still discouraging; and
(ii) conventional low-resolution ADC systems only reduce the number of quantization bits while still maintaining a large number of RF chains.
Consequently, an elaborate design of the low-resolution ADC receiver architecture is desirable to overcome such limitations. 
}


Recently, to reduce both the numbers of RF chains and quantization bits, hybrid analog-and-digital beamforming (HBF) is jointly considered with low-resolution ADCs~\cite{mo2017hybrid, abbas2017millimeter, roth2018comparison}.
HBF for low-resolution ADC systems achieves a potential EE gain compared to fully digital BF (DBF) with low-resolution ADC systems~\cite{mo2017hybrid}.
A naive combination of HBF with a low-resolution ADC receiver, however, cannot 
achieve the full potential of the large antenna arrays due to the increased quantization error~\cite{abbas2017millimeter,roth2018comparison}.
{\color{black} This shows the additional limitation of using a low-resolution ADC in terms of reducing the number of RF chains: state-of-the-art HBF techniques developed for perfect quantization systems cannot be directly used.} 
Hence, the HBF for low-resolution ADC receivers need to be re-designed to fully utilize its potential advantage.

In this article, we introduce advanced low-resolution ADC receiver architectures  for mmWave massive MIMO communications.
We first focus on the low-resolution ADC receiver architecture with HBF and present two advanced receiver architectures that further improve the SE-EE tradeoff by addressing key design challenges.
In the following sections, we detail the challenges in designing a HBF receiver with quantization error and highlight the potential design optimization aspects to manage the quantization error.
We further explain key insights on how the proposed advanced receiver architectures overcome such challenges based on both theoretical and numerical results.
In addition, we introduce an alternative receiver architecture, called a learning-based blind receiver, which does not require explicit channel estimation.
Finally, we conclude with a discussion on future research directions associated with remaining design challenges.

\section{Hybrid Beamforming for Low-Resolution ADC Receiver}
\label{sec:hybrid_BF_lowADCs}

HBF was initially introduced in \cite{zhang2005variable}.
The key idea is to project high-dimensional signals onto an low-dimensional subspace by employing analog BF (ABF) followed by processing using DBF \cite{alkhateeb2014channel,heath2016overview}.
HBF entered the limelight of standard organizations such as 3GPP, after mmWave communication was considered for 5G communications. 
The motivation for HBF in mmWave communications emerges from the sparse nature of the mmWave channels which enables analog/hybrid BF architectures to reduce the number of RF chains with negligible impact on the SE \cite{heath2016overview}.
To be specific, the HBF approach can be implemented in an energy-efficient manner for sparse mmWave channels by collecting most of the channel gains via ABF on a smaller number of RF chains and by managing inter-user (or intra-user) interference via DBF within the reduced dimensional subspace.
{\color{black}
Therefore, utilizing HBF for low-resolution ADC systems can create an opportunity to further improve the SE-EE tradeoff by reducing both the numbers of RF chains and quantization bits.}

\subsection{Advantages of Hybrid BF for Low-Resolution ADC Receiver}
\label{subsec:hybrid_BF}

By employing HBF, low-resolution ADC systems can further reduce the power consumption at the receivers thanks to the reduced number of RF chains, and hence can
be utilized in a variety of use cases such as drone communications and low-power-consumption Internet-of-Things device communications. 
Since such use cases require operation for a long period without a source of power supply, it is difficult to realize them with high-speed and high-resolution ADC systems.

In addition,
one of the pioneering works on HBF with low-resolution ADCs  showed that the HBF architecture achieves a better EE with low-resolution ADCs than with high-resolution ADCs \cite{mo2017hybrid}.
The simulation study in \cite{mo2017hybrid} further demonstrated that the HBF architecture can achieve a comparable SE to the fully-digital solution in the low and intermediate SNR range.
These results not only verify potential benefits of the HBF for low-resolution ADC systems with respect to EE but also confirm that its SE is not much degraded from that of the fully DBF for low-resolution ADC systems.

Related to these advantages, it is shown in \cite{abbas2017millimeter} that either HBF or fully DBF can achieve a better SE-EE tradeoff in low-resolution ADC systems depending on the power consumption characteristics of the mmWave receivers.
Although there are certain cases where the fully DBF yields the best SE-EE tradeoff, it is remarkable that for some other cases, the performance of HBF systems with even  simple hybrid combining methods can be comparable to or superior to that of fully DBF in the low-resolution ADC regime.

\subsection{ Challenges in Low-Resolution ADC System Design}

Although adopting HBF to low-resolution ADC systems is a potential power-efficient solution for massive MIMO communications, challenges are still remaining in optimizing the systems. 
Rigorous analysis for low-resolution ADC systems is complex since the quantization process is non-linear and the quantization error is intertwined with the channel coefficients.
To resolve the difficulty, there have been several linear quantization models used in the literature such as Bussgang decomposition and additive quantization noise model (AQNM).
Using the Bussgang decomposition, 
key communication functions such as a channel estimator were developed and analyized in one-bit ADC systems \cite{li2017channel}. 
Contrary to the Bussgang decomposition that often focuses on one-bit systems, the AQNM has been widely adopted in the literature to design or analyze general low-resolution ADC systems. 
Because of its tractability, low-resolution ADC systems were analyzed from both theoretical \cite{mo2017hybrid} and numerical perspectives~\cite{abbas2017millimeter}.
{\color{black} Regarding the HBF design,  it is still challenging to find the optimal BF coefficients in low-resolution ADC systems even with the linear quantization models.
This is because of the direct dependency between quantization error and the analog combined channels under the constant modulus constraint on each element of the analog combining matrix.
}

When designing HBF for low-resolution ADC systems, one should consider the main bottleneck in low-resolution ADC systems which is the limited resolution and dynamic range.
The quantization error becomes large in low-resolution ADC systems, and a small signal component can be easily buried in the quantization noise when the input signal power is high. 
Furthermore, a large input signal component could lead to ADC saturation, which can cause signal clipping. 
Those problems need to be addressed by re-designing the receiver's analog components such as 
1) analog combiner and 2) ADCs.
In the next section, we introduce novel architectures for mmWave communications with low-resolution ADCs.
\section{Advanced Receiver Architectures}
\label{sec:Advanced}

We introduce two advanced receiver architectures that re-design the analog components: the analog combiner and ADCs.
{\color{black} They employ phase-shifter based HBF for low-resolution ADC systems.}
Then, we briefly present an alternative data-driven receiver that decodes coarsely quantized signals without explicit channel knowledge.
{\color{black} 
The advanced receivers increase the SE while maintaining similar power consumption or even reducing it compared to conventional HBF for low-resolution ADC receiver architectures, thereby extending their SE-EE tradeoff region.
}
\subsection{Two-Stage Analog Combining Receiver}

\begin{figure}[!t]
    \centering
    \includegraphics[width=0.98\columnwidth]{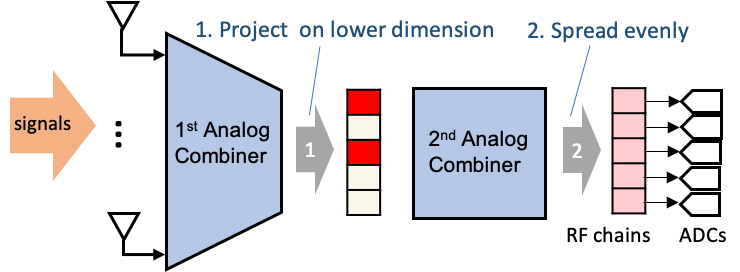}
    \caption{Two-stage analog combining architecture with key insights and principle.}
    \label{fig:twostage}
\end{figure}

{\color{black} Existing HBF techniques were developed for a high-resolution ADC which does not take into account the impact of the non-negligible quantization error. 
Accordingly, a two-stage analog combining receiver architecture was proposed for mmWave communications in~\cite{choi2019two} as a near optimal analog combining architecture for low-resolution ADC systems  in maximizing capacity.}
As shown in Fig.~\ref{fig:twostage}, the two-stage analog combining receiver employs two consecutive analog combiners with different sizes.
For the first analog combiner, the number of outputs is less than the number of receive antennas $N_r$ and equals to the number of RF chains $N_{\rm RF}$.
The numbers of inputs and outputs of the second analog combiner are the same as $N_{\rm RF}$, i.e., an $N_{\rm RF}$ square matrix.

Asymptotic optimality in the number of RF chains can be achieved if (1) the first analog combiner is the left-singular vectors of the channel matrix and (2) the second analog combiner is any $N_{\rm RF}$ dimensional orthonormal matrix that satisfies the constant modulus condition on its elements \cite{choi2019two}.
When the optimality conditions are met, the capacity increases logarithmically with the number of RF chains, which is an optimal scaling law, whereas the capacity with only the first analog combiner is upper bounded by a constant value. 
In addition, the capacity with the two-stage analog combiner can be maximized when nonzero channel singular values are the same.

Fig.~\ref{fig:twostage} shows the insight on how the two-stage analog combiner achieves such optimality.
When the described conditions are met, the first analog combiner collects the entire channel gains into the smaller dimension and the second analog combiner spreads the gains evenly over all RF chains.
Consequently, there is ideally no loss in the channel gain, and all ADCs generate moderate quantization errors thanks to the balanced signal power over the RF chains. 
Although the second analog combiner matrix does not reduce the total quantization error since it is orthonormal, it plays a key role in reducing the quantization error for the desired signals, {\color{black} which we define as {\it effective quantization error},} by changing the distribution of the channel gains.

In \cite{choi2019two}, the first analog combiner is designed with array response vectors (ARVs) to satisfy the constant modulus condition while achieving the performance of singular vector decomposition combining.
Exploiting the sparse mmWave channels in the beam domain, the ARVs can nearly meet the optimality condition for the first analog combiner.
The second analog combiner is designed with a discrete Fourier transform (DFT) matrix or Hadamard matrix  which adds {\color{black} the quadratic computational complexity of $N_{\rm RF}$}.
This shows that the two-stage analog combining architecture provides a favorable structure for implementing the analog combiner that closely satisfies the optimality conditions with a constant modulus constraint on each element.
In addition, the second analog combiner does not depend on the channel conditions.
Therefore, it can be implemented with fixed phase shifters without requiring additional power consumption on steering the phases.
Furthermore, if $N_{\rm RF}$ is a power of two, the fast Fourier transform version of the DFT calculation can be implemented, which only adds {\color{black} the linearithmic complexity of $N_{\rm RF}$}
 compared to the conventional HBF.

\begin{figure}[!t]
    \centering
    \includegraphics[width=1\columnwidth]{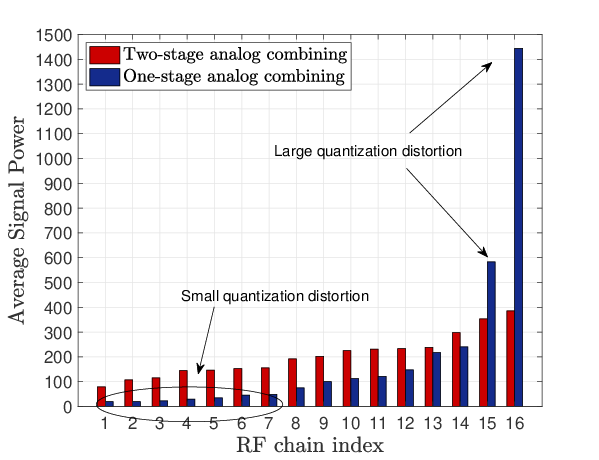}
    \caption{The average signal power on each RF chain for $128$ BS antennas, $16$ RF chains, $4$ users, $64$-QAM constellation, and $2$ average channel paths at ${\rm SNR} = 10$ dB.}
    \label{fig:sigpow}
\end{figure}


Adopting a narrowband geometry-based channel model as in \cite{choi2019two} for $2$ average channel paths, the average signal powers on each RF chain are compared in Fig.~\ref{fig:sigpow} for the two-stage and conventional one-stage analog combining cases when the first analog combiner is designed with ARVs as in \cite{choi2019two}.
The signal power of the two-stage case is distributed evenly over the RF chains whereas that of the one-stage  case is concentrated on a few RF chains.
Accordingly, for the one-stage case, the large signals on a few RF chains, which are mostly desired signal components, suffer from large quantization distortion while the small signals on many of the RF chains, which are mostly noise components, experience small quantization distortion. 
Consequently, adding the second analog combiner can reduce the effective quantization error by evenly spreading the signal power. 
Thus, it potentially mitigates the forementioned problems caused by the  limited resolution and dynamic range of ADCs.

\subsection{Resolution-Adaptive ADC Receiver}

\begin{figure}[!t]
    \centering
    \includegraphics[width=0.93\columnwidth]{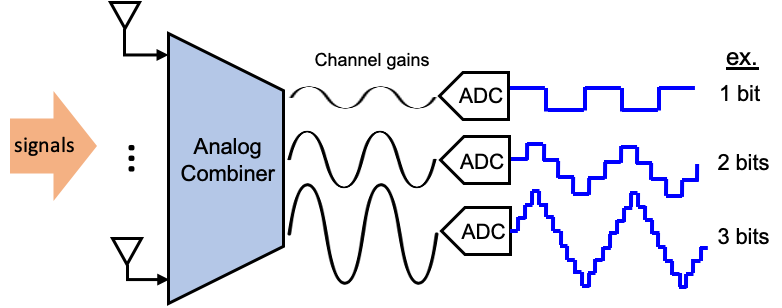}
    \caption{Resolution-adaptive ADC with HBF architecture with key insights and principle.}
    \label{fig:adaptive}
\end{figure}

{\color{black} To apply the existing state-of-the-art HBF techniques developed for perfect quantization systems with minimal performance degradation, a resolution-adaptive ADC receiver was proposed in~\cite{choi2017resolution}.
This work provides a near optimal ADC bit distribution that minimizes the total mean square quantization error (MSQE) under limited power consumption, which leads to  an increase in the SE.}
As shown in Fig.~\ref{fig:adaptive}, the proposed receiver changes the resolution of each ADC depending on the aggregated channel gain on each ADC, thereby using a limited power budget efficiently. 
{\color{black} To this end, the receiver first computes the closed-form solution in~\cite{choi2017resolution} based on a channel estimate to determine bit allocation, and use the bit allocation to change the ADC bit resolution. }

The number of allocated bits for each ADC increases logarithmically with the associated channel gain, i.e., more bits to the ADC with a larger channel gain as shown in Fig.~\ref{fig:adaptive}.
The derived bit-allocation solution has two different interpretations: (1) in the high signal-to-noise ratio (SNR) range, it minimizes the total quantization error  and (2) in the low SNR range, it maximizes the generalized mutual information which serves as a lower bound on channel capacity.
Therefore, the derived bit allocation solution is expected to be effective in both high and low SNR ranges.

When the power consumption of $N_{\rm RF}$ ADCs with a fixed resolution of $\bar b$ bits is set as a power constraint, {\color{black} the near optimal bit allocation for the $i$th pair of ADCs becomes logarithmically proportional to the cube root of the aggregated channel gain at the $i$th RF chain.}
Once the bit distribution is determined, it remains unchanged during the channel coherence time or a longer time duration such as the coherence time for the large scale fading or the angles of arrivals or departures.
 Since the bit allocation is a closed-form solution \cite{choi2017resolution}, the additional computational complexity is {\color{black} the linear of $N_{\rm RF}$ and $N_u$} for computing the aggregated channel gains where $N_u$ is the number of transmitted symbols. 

\begin{figure}[!t]
    \centering
    \includegraphics[width=0.95\columnwidth]{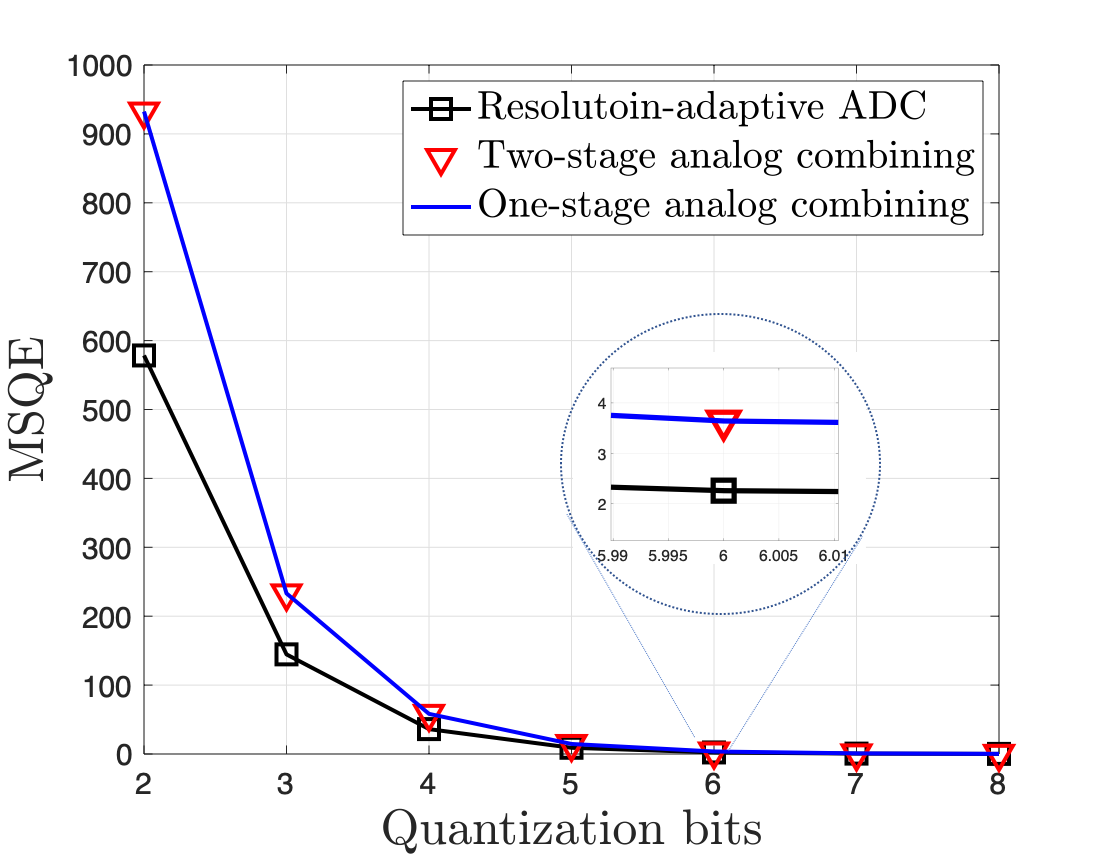}
    \caption{The total MSQE  for $128$ BS antennas, $16$ RF chains, $4$ users, $64$-QAM constellation, and $2$ average channel paths at ${\rm SNR} = 10$ dB. 
    Quantization bits indicate the constraint bits $\bar b$ for the adaptive ADC case and the actual ADC bits for the other cases.}
    \label{fig:msqe}
\end{figure}

Since the bit allocation solution minimizes the total quantization errors, this also reduces the effective quantization error. 
Therefore, it can also mitigate the problems caused by the limited resolution and dynamic range of low-resolution ADCs.
Note that the beam domain channel vector that the receiver sees after the analog combining is sparse in mmWave communications.
Accordingly, more selective bit allocations can be realized, thereby providing a larger benefit.
In Fig.~\ref{fig:msqe}, the total MSQE results with the same ARV-based analog combiner demonstrate that the adaptive ADC receiver reduces the quantization error more than 30\% from the other receivers with fixed-resolution ADCs.
Therefore, the state-of-the-art HBF techniques for perfect quantization systems can be applied with reduced performance degradation.


{\color{black} As an example of the bit allocation result, $27.5\%$, $47.7\%$, $23.54\%$, and $1.25\%$ of the ADCs are allocated 0, 1, 2, and 3 bits on average, respectively, for the same system configuration as Fig.~\ref{fig:msqe} with the constraint bit $\bar b =1$}.
This means that allocating equal bits to all ADCs is far from minimizing the quantization error. 
In addition, when the bit constraint is very small, some ADCs are assigned with zero bit.
The RF components associated with zero-bit ADCs can be turned off during the channel coherence period, thereby providing additional power saving to the receiver.


\subsection{Learning-based Blind Receiver}

\begin{figure}[!t]
    \centering
    \includegraphics[width=0.95\columnwidth]{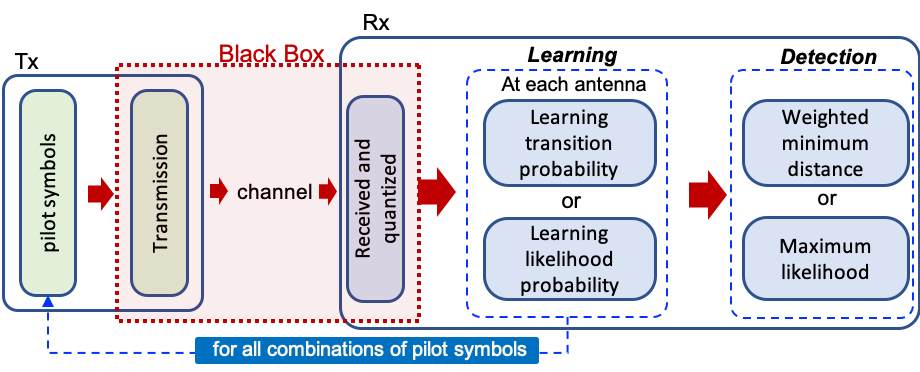}
    \caption{Learning based blind detection receiver architecture with key insights and principle.}
    \label{fig:learning}
\end{figure}

Here we introduce a different type of advanced receivers as a guide toward future architectures.
{\color{black} To avoid channel estimation which is particularly more challenging in low-resolution ADC systems, a learning-based blind receiver architecture was developed in~\cite{hong2017weighted,choi2019robust}.
This receiver decodes quantized signals without requiring any explicit channel estimation and provides similar performance to the receiver with perfect channel information.}
As shown in Fig.~\ref{fig:learning}, the fundamental idea behind this receiver is to learn characteristics of the quantized outputs at each antenna instead of estimating the channel state information (CSI), by considering the channel and quantization functions as a black box.
When the quantized outputs are obtained during the training phase, the receiver counts the frequency of each quantized output and considers it as a likelihood function for an associated training symbol.
Leveraging the likelihood functions, the receiver can perform symbol detection using, for example, the maximum likelihood and minimum distance detection.

However, the main problem of this receiver is the large dependency on the training length for each combination of transmit symbols $N_{\rm tr}$. 
Particularly in the medium to high SNR regime, it is highly likely to obtain the same quantized outputs at many antennas during the training phase, i.e., $N_{\rm tr}$ zeros or ones at each antenna for the one-bit ADC case, since the channel is invariant during the training.
Then, the likelihood probability for the training symbol at such antenna is either zero or one.
This does not provide valid information when using maximum likelihood detection, and even wipes out the entire information obtained during the training phase, thereby severely degrading the detection capability.
Therefore, a prohibitively large number of training symbols is required to obtain desirable likelihood functions over most antennas at every channel coherence time.

A coding theoretic approach \cite{hong2017weighted} and a signal processing approach \cite{choi2019robust} depicted in Fig.~\ref{fig:learning} were proposed to overcome such challenge, by significantly reducing the training length $N_{\rm tr}$.
In~\cite{hong2017weighted}, the receiver first  estimates the quantized outputs without additive noise 
and considers the estimated outputs as a codeword. 
Then, the transition probability of each element of the codeword is also estimated at each antenna.
Using the estimated codeword and transition probabilities, the receiver decodes the transmitted symbols based on the weighted minimum distance detection. 
As a different approach, a dithering technique is used during the training phase in \cite{choi2019robust} to increase the robustness of learning likelihood functions.  
Adding dithering signals provides artificial noise to the received signals and decreases the probability of observing $N_{\rm tr}$ zeros or ones. 
Then, the actual likelihood function is derived from the obtained likelihood functions of the dithered signals and used for maximum likelihood detection.

Although the proposed techniques greatly reduce the training length $N_{\rm tr}$ for each training symbol, {\color{black} the number of training symbols that increases exponentially with the number of users still remains as a primary bottleneck of realizing such receiver, where $M$ and $N_u$ denote the modulation order and the number of user symbols.}
Therefore, this architecture still needs more investigation to manage the large number of training symbols $N_{\rm sym}$ when there are many users with high-order modulations.
Despite of this limitation, the learning-based blind receiver can be considered as a potential receiver architecture especially for one-bit ADC systems since channel estimation is particularly more challenging in one-bit ADC systems.
{\color{black} In addition, switch-based analog combining can be further adopted to reduce the number of RF chains by selecting only the antennas that provide well-trained likelihood functions.}

\section{Spectral and Energy Efficiency}
\label{sec:SE_EE}

\begin{figure}[!t]
    \centering
    \includegraphics[width=1\columnwidth]{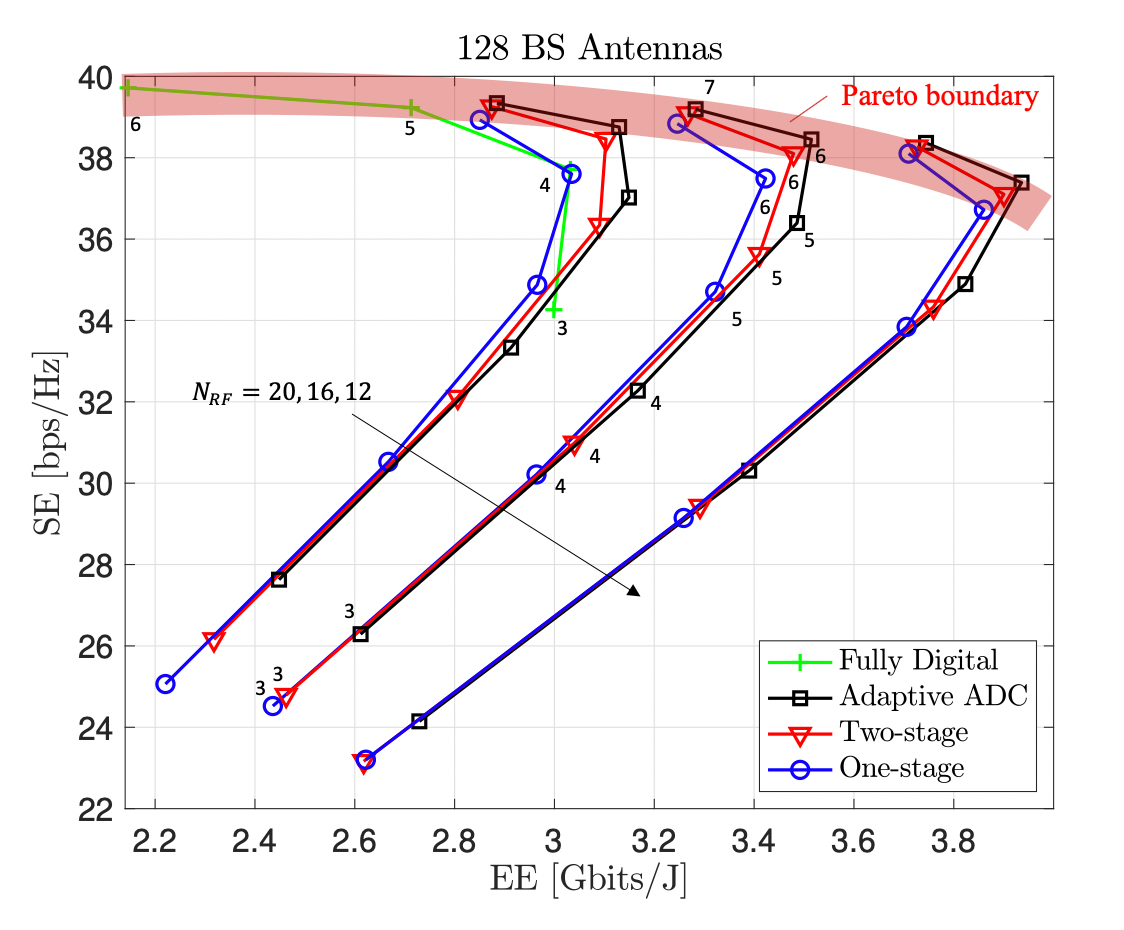}
    \caption{Simulation results with $4$ users, $2$ average channel paths,  ${\rm SNR} = 10$ dB, and $f_s = 1$ GHz bandwidth for $\{12, 16, 20\}$ RF chains.
    SE and EE points for $3-7\ {\rm bits}$ are simulated.
    The Pareto boundary is for each method in the numbers of bits and RF chains.}    
    \label{fig:SE_EE}
\end{figure}

{\color{black} Since maximizing SE would hurt EE and vice versa, achieving the best SE-EE tradeoff is important.}
In Fig.~\ref{fig:SE_EE}, the SE-EE tradeoff is evaluated for the two-stage analog combining, resolution-adaptive ADC, and one-stage analog combining receivers with the channel model used in \cite{choi2019two}.
ARV-based analog combiners are used for the first analog combiner of the two-stage analog combining receiver as in \cite{choi2019two} and also used for analog combiners of the resolution-adaptive ADC and one-stage analog combining receivers.
Each point in the curves represents the SE-EE tradeoff result for $3$ to $7$ quantization bits per ADC for the case of $128$ base station (BS) antennas.
In computing the power consumption of analog components, we use $P_{\rm LNA} = 39$mW, $P_{\rm PS} = 2$mW, $P_{\rm M}=16.8$mW, $P_{\rm LO}=5$mW, $P_{\rm LPF}=14$mW, and $P_{\rm BBamp}=5$mW  for a low noise amplifier, phase shifter, mixer, local oscillator, low-pass filter, baseband amplifier \cite{abbas2017millimeter}, respectively
We use  $P_{\rm ADC}$ in \cite{choi2017resolution} for the ADC power consumption.

Unlike the fully DBF with low-resolution ADC receiver, the HBF with low-resolution ADC receivers provide various SE-EE tradeoffs for both different numbers of RF chains and bits with an extended SE-EE tradeoff region. 
{\color{black} As the number of bits increases, the SE and EE both increase, and the EE starts to decrease due to the larger increase in power consumption compared to the increase in the SE. 
When the number of RF chains increases, although the SE increases as more channel gains can be collected, the EE can decrease if the SE increase is smaller than the increase in power consumption.
Therefore, there exists an optimal tradeoff between the numbers of RF chains and bits.}
The HBF based receivers achieve the best SE-EE tradeoff with $6$ to $7$ bits per ADC for $12$ to $20$ RF chains, whereas the fully DBF based receiver shows the best tradeoff with $4$ to $5$ bits and suffers from high power consumption for more than $5$ bits due to the large number of RF chains and ADCs.
{\color{black} The best SE-EE tradeoffs achieved by the HBF-based low-resolution ADC receivers can have both high SE and EE whereas the fully digital low-resolution ADC receiver shows limited EE performance with high SE.}
This demonstrates the advantage of deploying the HBF for low-resolution ADC systems.

Among the HBF based receivers, the resolution-adaptive ADC receiver shows the best SE-EE performance since it reduces both the total quantization error and effective quantization error. 
Although the two-stage analog combining receiver does not reduce the total quantization error, it also shows an improved SE-EE performance compared to the one-stage analog combining receiver by reducing the effective quantization error.
We remark that for the same number of ADC bits, the adaptive ADC receiver achieves the highest SE and the two-stage analog combining receiver reveals an improved SE in most cases compared to the conventional one-stage analog combining receiver.
Thus, the two advanced receivers achieve higher SE-EE tradeoffs.

\section{Key Challenges and Future Directions}
\label{sec:extension}

In this section, we highlight the key challenges and potential research directions for mmWave receiver architectures with low-resolution ADCs.

\textbf{Extensions to sub-array architectures:} The current 3GPP standards of 5G NR are mainly developing under a sub-array HBF architecture because of its relatively simple implementation. 
It is therefore important to extend the proposed low-resolution ADC receivers to the sub-array architecture where only a subset of antennas is connected to each RF chain.
For example, one initial approach would be to use a sub-array analog combiner as a first analog combiner followed by a second analog combiner that satisfies the constant modulus condition. 
Based on the same insight shown in Fig.~\ref{fig:twostage}, one can expect performance improvement from a conventional sub-array based HBF receiver with low-resolution ADCs.

\textbf{Extensions to frequency-selective channels:} Another important research direction is to identify potential advantages of the proposed architectures in an orthogonal frequency-division multiplexing system. 
The advanced receivers have been analyzed under the narrowband channel assumption to validate their advantages in the initial stage.
Nonetheless, assuming that the first stage analog combiner is designed to match strong signal paths in a wideband sense as the standards specify, it is more likely for the second stage analog combiner to play a key role in reducing the effective quantization errors in the mmWave bands.
Developing efficient solutions to address these challenges in wideband channels is an important future research direction.

\textbf{Channel estimation:} 
In the 5G standards, ABF of the transceiver is selected from pre-defined codebooks through beam sweeping.
Then, the channel estimation is performed via sounding reference signals to obtain the CSI of the effective channels after composing the ABF.
Therefore, it is somewhat challenging, in the current standards, to select the ABF in a sophisticated manner since it should be done without the CSI.
For the proposed architectures, the average signal-to-quantization-noise-ratio of the pilot symbols could be improved from the conventional architecture.
Accordingly, considering the various aspects of the 3GPP standards, it is important to develop efficient channel estimation strategies for mmWave receivers with low-resolution ADCs. These strategies could also leverage prior observations and machine learning tools to reduce the channel training and estimation overhead \cite{zhang2019}. 


\section{Conclusion}
\label{sec:con}

Low-resolution ADCs provide advantages in reducing hardware cost and power consumption in mmWave massive MIMO systems. 
However, the performance degradation due to non-negligible quantization error and the large number of RF chains are still not desirable.
In this article, we described advanced receiver architectures that employ low-resolution ADCs and described key insights and underlying principles.
Employing hybrid beamforming jointly with low-resolution ADCs offers the opportunity to reduce both the number of RF chains and quantization bits. 
Leveraging this benefit, the two-stage analog combining and resolution-adaptive ADC receivers focus on optimizing the analog combiner and ADC resolutions to reduce the effective quantization error, respectively.
We also introduced a learning-based low-resolution ADC receiver which decodes symbols without the CSI via a simple learning process and can be considered as a potential data-driven receiver architecture for one-bit ADC systems.
Although many challenges still remain to realize the proposed receivers, theoretical and numerical studies validated their high potential.

\bibliographystyle{IEEEtran}
\bibliography{ComMag.bib}

\begin{thebibliography}{10}
\providecommand{\url}[1]{#1}
\csname url@samestyle\endcsname
\providecommand{\newblock}{\relax}
\providecommand{\bibinfo}[2]{#2}
\providecommand{\BIBentrySTDinterwordspacing}{\spaceskip=0pt\relax}
\providecommand{\BIBentryALTinterwordstretchfactor}{4}
\providecommand{\BIBentryALTinterwordspacing}{\spaceskip=\fontdimen2\font plus
\BIBentryALTinterwordstretchfactor\fontdimen3\font minus
  \fontdimen4\font\relax}
\providecommand{\BIBforeignlanguage}[2]{{%
\expandafter\ifx\csname l@#1\endcsname\relax
\typeout{** WARNING: IEEEtran.bst: No hyphenation pattern has been}%
\typeout{** loaded for the language `#1'. Using the pattern for}%
\typeout{** the default language instead.}%
\else
\language=\csname l@#1\endcsname
\fi
#2}}
\providecommand{\BIBdecl}{\relax}
\BIBdecl

\bibitem{pi2011introduction}
Z.~Pi and F.~Khan, ``{An introduction to millimeter-wave mobile broadband
  systems},'' \emph{IEEE Commun. Mag.}, vol.~49, no.~6, pp. 101--107, 2011.

\bibitem{zhang2018low}
J.~Zhang, L.~Dai, X.~Li, Y.~Liu, and L.~Hanzo, ``{On low-resolution ADCs in
  practical 5G millimeter-wave massive MIMO systems},'' \emph{IEEE Commun.
  Mag.}, vol.~56, no.~7, pp. 205--211, 2018.

\bibitem{liu2019low}
J.~Liu, Z.~Luo, and X.~Xiong, ``{Low-Resolution ADCs for Wireless
  Communication: A Comprehensive Survey},'' \emph{IEEE Access}, vol.~7, pp.
  91\,291--91\,324, 2019.

\bibitem{mo2017hybrid}
J.~Mo, A.~Alkhateeb, S.~Abu-Surra, and R.~W. Heath, ``{Hybrid architectures
  with few-bit ADC receivers: Achievable rates and energy-rate tradeoffs},''
  \emph{IEEE Trans. on Wireless Commun.}, vol.~16, no.~4, pp. 2274--2287, 2017.

\bibitem{abbas2017millimeter}
W.~B. Abbas, F.~Gomez-Cuba, and M.~Zorzi, ``{Millimeter wave receiver
  efficiency: A comprehensive comparison of beamforming schemes with low
  resolution ADCs},'' \emph{IEEE Trans. on Wireless Commun.}, vol.~16, no.~12,
  pp. 8131--8146, 2017.

\bibitem{roth2018comparison}
K.~Roth, H.~Pirzadeh, A.~L. Swindlehurst, and J.~A. Nossek, ``{A comparison of
  hybrid beamforming and digital beamforming with low-resolution ADCs for
  multiple users and imperfect CSI},'' \emph{IEEE J. of Sel. Topics in Signal
  Process.}, vol.~12, no.~3, pp. 484--498, 2018.

\bibitem{zhang2005variable}
X.~Zhang, A.~F. Molisch, and S.-Y. Kung, ``{Variable-phase-shift-based
  RF-baseband codesign for MIMO antenna selection},'' \emph{IEEE Trans. on
  Signal Process.}, vol.~53, no.~11, p. 4091, 2005.

\bibitem{alkhateeb2014channel}
A.~Alkhateeb, O.~El~Ayach, G.~Leus, and R.~W. Heath, ``{Channel estimation and
  hybrid precoding for millimeter wave cellular systems},'' \emph{IEEE J. of
  Sel. Topics in Signal Process.}, vol.~8, no.~5, pp. 831--846, 2014.

\bibitem{heath2016overview}
R.~W. Heath, N.~Gonzalez-Prelcic, S.~Rangan, W.~Roh, and A.~M. Sayeed, ``{An
  overview of signal processing techniques for millimeter wave MIMO systems},''
  \emph{IEEE J. of Sel. Topics in Signal Process.}, vol.~10, no.~3, pp.
  436--453, 2016.

\bibitem{li2017channel}
Y.~Li, C.~Tao, G.~Seco-Granados, A.~Mezghani, A.~L. Swindlehurst, and L.~Liu,
  ``{Channel estimation and performance analysis of one-bit massive MIMO
  systems},'' \emph{IEEE Trans. on Signal Process.}, vol.~65, no.~15, pp.
  4075--4089, 2017.

\bibitem{choi2019two}
J.~Choi, G.~Lee, and B.~L. Evans, ``{Two-Stage Analog Combining in Hybrid
  Beamforming Systems With Low-Resolution ADCs},'' \emph{IEEE Trans. on Signal
  Process.}, vol.~67, no.~9, pp. 2410--2425, 2019.

\bibitem{choi2017resolution}
J.~Choi, B.~L. Evans, and A.~Gatherer, ``{Resolution-adaptive hybrid MIMO
  architectures for millimeter wave communications},'' \emph{IEEE Trans. on
  Signal Process.}, vol.~65, no.~23, pp. 6201--6216, 2017.

\bibitem{hong2017weighted}
S.-N. Hong, S.~Kim, and N.~Lee, ``{A weighted minimum distance decoding for
  uplink multiuser MIMO systems with low-resolution ADCs},'' \emph{IEEE Trans.
  on Commun.}, vol.~66, no.~5, pp. 1912--1924, 2017.

\bibitem{choi2019robust}
J.~Choi, Y.~Cho, B.~L. Evans, and A.~Gatherer, ``{Robust Learning-Based ML
  Detection for Massive MIMO Systems with One-Bit Quantized Signals},''
  \emph{IEEE Global Commun. Conf.}, Dec. 9-13, 2019.

\bibitem{zhang2019}
{Y. Zhang, M. Alrabeiah, and A. Alkhateeb}, ``{Deep Learning for Massive MIMO
  with 1- Bit ADCs: When More Antennas Need Fewer Pilots},'' \emph{arXiv
  preprint arXiv:1910.06960}, 2019.

\end{thebibliography}
\newpage

\begin{IEEEbiographynophoto}
{Jinseok Choi} [M'19] (jinseokchoi89@utexas.edu) received his B.S. degree in Electrical and Electronic Engineering at Yonsei University, Seoul, South Korea in 2014. He received his M.S. and Ph.D. degrees in Electrical and Computer Engineering at The University of Texas at Austin, TX, USA, in 2016 and 2019, respectively. 
He is now a wireless R\&D senior engineer at Qualcomm Inc., San Diego, CA, USA.
\end{IEEEbiographynophoto}

\begin{IEEEbiographynophoto}
{Gilwon Lee} [M'16] (gilwon.lee30@gmail.com) is a staff research engineer at Samsung Research America (SRA), Plano, TX. He earned his Ph.D. degree in Electrical Engineering from Korea Advanced Institute of Science and Technology (KAIST) in 2016. He was a postdoctoral fellow in Electrical and Computer Engineering at The University of Texas at Austin, and was a software engineer at Intel Corporation, Santa Clara, CA. 
\end{IEEEbiographynophoto}

\begin{IEEEbiographynophoto}
{Ahmed Alkhateeb} [M'16] (aalkhateeb@asu.edu) received the B.S. and M.S. degrees in electrical engineering from Cairo University, Egypt, in 2008 and 2012, respectively, and the Ph.D. degree in electrical engineering from The University of Texas at Austin, USA, in 2016. From 2016 to 2017, he was a Wireless Communications Researcher with the Connectivity Lab, Facebook. 
He joined Arizona State University in 2018, where he is currently an Assistant Professor with the School of Electrical, Computer and Energy Engineering. 
He was a recipient of the 2016 IEEE Signal Processing Society Young Author Best Paper Award.
\end{IEEEbiographynophoto}

\begin{IEEEbiographynophoto}
{Alan Gatherer}[F'15] (alan.gatherer@futurewei.com) is a Senior Technical Vice President at Futurewei and an IEEE Fellow.  He is responsible for software/hardware R\&D for next generation baseband SoC. Alan joined Futurewei in January 2010. Previously he was a Fellow and CTO at Texas Instruments leading the development of high performance, multicore DSP and 3GPP standardization. Alan has authored over 50 journal and conference papers, holds over 80 awarded patents and is author of "The Application of Programmable DSPs in Mobile Communications". Alan received his BEng from Strathclyde University in Scotland and his PhD from Stanford University in California in 1993. 

\end{IEEEbiographynophoto}

\begin{IEEEbiographynophoto}
{Naofal Al-Dhahir}[F'07] (aldhahir@utdallas.edu) is Erik Jonsson Distinguished Professor and ECE Associate Head at UT-Dallas. He earned his PhD degree in Electrical Engineering from Stanford University.  From 1994 to 2003, he was a principal member of the technical staff at GE Research and AT\&T Shannon Laboratory.  
He is co-inventor of 43 issued US patents, co-author of over 430 papers and co-recipient of 4 IEEE best paper awards. He received the 2019 IEEE SPCE technical recognition award. He is an IEEE Fellow and served as Editor-in-Chief of IEEE Transactions on Communications from 2016 to 2019.
\end{IEEEbiographynophoto}

\begin{IEEEbiographynophoto}
{Brian L. Evans} [M'87-SM'97-F'09] (bevans@ece.utexas.edu) received the B.S. degree in electrical engineering and computer science from the Rose-Hulman Institute of Technology in 1987, and M.S. and Ph.D. degrees in electrical engineering from Georgia Tech in 1988 and 1993. After three years as a post-doctoral researcher at the University of California, Berkeley, he joined the faculty at The University of Texas at Austin. His group develops signal processing algorithms with implementation constraints in mind, and translates algorithms into design methods and embedded prototypes. He has published 270 papers and graduated 29 Ph.D. students. He received a 1997 U.S. NSF CAREER Award.

\end{IEEEbiographynophoto}

\end{document}